\documentclass[12pt]{article}
\usepackage{graphicx}
\usepackage{amssymb}
\usepackage{amsmath}
\usepackage{color}

\usepackage{hyperref}

\begin{document}

\begin{center}
{\Large \textbf{Particle decay, Oberth effect and
a relativistic\\[7pt]
rocket in the Schwarzschild background}}

\

\textbf{Yu. V. Pavlov$^{1,2}$} and \textbf{O. B. Zaslavskii$^{3}$}\\[11pt]

${}^{1}$\,Institute of Problems in Mechanical Engineering, Russian Academy
of Sciences,\\[0pt]
61~Bol'shoy pr., St. Petersburg 199178, Russia;\\[0pt]
${}^{2}$\,N.I.\,Lobachevsky Institute of Mathematics and Mechanics, Kazan
Federal University, 18 Kremlyovskaya St., Kazan 420008, Russia;\\[0pt]
E-mail: yuri.pavlov@mail.ru

${}^{3}$\,Department of Physics and Technology, Kharkov V.N.\,Karazin
National University, 4~Svoboda Square, Kharkov 61022, Ukraine\\[0pt]
E-mail: zaslav@ukr.net
\end{center}

\begin{abstract}
We relate the known Oberth effect and the nonrelativistic analogue of the
Penrose process. When a particle decays to two fragments, we derive the
conditions on the angles under which debris can come out for such a process
to occur. We also consider the decay and the Oberth effect in the
relativistic case, when a particle moves in the background of the
Schwarzschild black hole. This models the process when a rocket ejects fuel.
Different scenarios are analyzed depending on what data are fixed. The
efficiency of the process is found, in particular near the horizon and for a
photon rocket (when the ejected particle is massless). We prove directly
that the most efficient process occurs when fuel is ejected along the rocket
trajectory. When this occurs on the horizon, the efficiency reaches
100\thinspace \% for a photon rocket. We compare two ways how a rocket can
reverse its direction of motion to a black hole near the event horizon with
restoring the initial energy-to-mass ratio: (i) by a single ejection or (ii)
in the two-step process when it stops and moves back afterwards. For a
nonphotonic rocket, in case (ii) a larger mass can be taken out from the
vicinity of a horizon. For a photonic one, there is no difference between
(i) and (ii) in this respect. We also consider briefly the scenario when a
rocket hangs over a black hole due to continuous ejection of fuel. Then, the
fuel mass decays exponentially with the proper time.
\end{abstract}


%% 04.20.-q Classical general relativity
%% 04.20.Cv Fundamental problems and general formalism
%% 04.70.Bw Classical black holes

%%%%%%%%%%%%%%%%%%%%%%%%%%%%%%%%%%%%%%%%%%%%%%%%%%%%%%%%%%%%%%%%%%%%%%

\section{Introduction}

In 1929, Oberth pointed out that using of the reactive fuel becomes more
efficient for rapid motion of a rocket~\cite{Oberth1929}.
Its engine performs more useful work than in the case of slow motion.
Indeed, in the laboratory frame, fuel flowing out from a nozzle of a moving
rocket can have a velocity less than that of a jet stream from a slow moving
rocket. As a consequence, the kinetic energy of jet fuel can decrease,
whereas the energy used for rocket acceleration can increase.

In the gravitational field, high speed is achieved in a lowest point of the
trajectory. Therefore, the use of jet fuel in periastron is more efficient
than in highest points. In the nonrelativistic case, this can be explained
on the basis of the energy conservation law since the total energy of jet
fuel is the sum of positive kinetic and negative potential energies of fuel
in the gravitational field.
    (In the nonrelativistic case, a reader can find simple pedagogical
presentation of the Oberth effect in a methodical paper~\cite{bm}).

    If the sum of the potential and kinetic energies of used jet fuel is
negative in the gravitational field, it was firstly noticed
in~\cite{GribPv2020} that the Oberth effect can be thought of
as a nonrelativistic realization of the effect similar to the Penrose one
in the ergosphere of a black hole~\cite{Penrose69}, \cite{PenroseFloyd71}.
    In this case, the total
nonrelativistic energy of a rocket after using jet fuel becomes bigger than
the energy of a rocket plus fuel before turning on an engine.

    In the present paper, we consider the Oberth effect for a relativistic case
when a rocket moves in the metric of a nonrotating black hole. But we start
with the short discussion of a nonrelativistic case.

    It is worth noting that general approach to the motion of a rocket in a
general relativity was developed in~\cite{HN12}, and considered for the
Schwarzschild metric in~\cite{Nat}. There, the main accent was made on a
motion of a body as such. Meanwhile, our main goal is to include the issue
under consideration in a context connected with particle decay. In doing so,
special attention is paid to the processes in the vicinity of a black hole
horizon.

    There is also one more line of motivation. Recently, motion of an
accelerated observer that moves in the Schwarzschild background attracted
some attention~\cite{new1,new2}.
    In particular, this opens one more~\cite{new3} scenario of the
so-called Ba\~{n}ados-Silk-West effect~\cite{new4}.
In the aforementioned papers only kinematics of motion was considered.
Meanwhile, we are interested in dynamic process that occur with a moving
observer (rocket).
    A real rocket ejects fuel continuously.
    However, to simplify matter, we model this process by considering breakup
of particle to another two. Continuous ejection is considered
in a separate Section~16.

%%%%%%%%%%%%%%%%%%%%%%%%%%%%%%%%%%%%%%%%%%%%%%%%%%%%%%%%%%%%%%%%%%%%%

\section{Nonrelativistic Case}

    Let a point-like particle of the mass $m_0$ decay to two particles at a
distance $r$ from a large attracting mass $M$ ($M\gg m_{0}$).
    We denote $ m_{1} $ and $m_{2}$ masses of debris.
    We consider process of ejection of fuel
under an arbitrary angle to the direction of motion of a rocket but, with
the restriction, that all particles move within the same plane.

Let us consider decay of particle 0 to 1 and 2 in point $r$. Particle~0
moves with the velocity $\mathbf{v}_{b}$, particles 1, 2 move with the
velocities $\mathbf{v}_{1,2}$. In the present paper, particle~2 corresponds to
a rocket, particle~1 corresponds to fuel. We have for the energy
    \begin{equation}
E_{0}=\frac{m_{0}}{2} {v}_{b}^{2}-\frac{Gm_{0}M}{r} ,  \label{e0g}
\end{equation}
$G$ is the gravitational constant, $E_{0}$ is the total mechanical energy of
an original body. If $E_{0}\geq 0$, it can move along an unbound trajectory,
then
\begin{equation}
E_{0}=\frac{m_{0} {v}_{0}^{2}}{2},  \label{e0}
\end{equation}
    where $\mathbf{v}_{0}$ is the velocity at infinity. Thus
    \begin{equation}
{v}_{b}^{2}= {v}_{0}^{2}+2G\frac{M}{r}.  \label{vb}
\end{equation}
The total mass
\begin{equation}
m_{0}=m_{1}+m_{2} .  \label{summ}
\end{equation}
Let superscript (0) denote quantities calculated in the frame comoving with
particle~0. This frame corresponds to the center of mass (CM) of particles~1
and~2. Then, the relative velocity $\mathbf{u}$ between particles~1 and~2
equals
\begin{equation}
\mathbf{u}=\mathbf{v}_{2}-\mathbf{v}_{1}=\mathbf{v}_{2}^{(0)}-\mathbf{v}_{1}^{(0)} .
\label{uot}
\end{equation}
From~(\ref{uot}) and the conservation law for the momentum we have
\begin{equation}
\mathbf{v}_{1}=\mathbf{v}_{b}-\frac{m_{2}}{m_{0}}\mathbf{u},  \label{v1b}
\end{equation}
\begin{equation}
\mathbf{v}_{2}=\mathbf{v}_{b}+\frac{m_{1}}{m_{0}}\mathbf{u}.  \label{v2b}
\end{equation}
From~(\ref{e0g}), (\ref{summ}), (\ref{v1b}), (\ref{v2b}) we get
\begin{equation}
E_{1}+E_{2}=E_{0}+E_{f},  \label{UnTE}
\end{equation}
    where
\begin{equation}
E_{1}=m_{1}\left( \frac{ {v}_{1}^{2}}{2}-\frac{GM}{r}\right) , \ \ \ \
E_{2}=m_{2}\left( \frac{ {v}_{2}^{2}}{2}-\frac{GM}{r}\right) ,  \label{UnTE2}
\end{equation}
and the quantity $E_{f}$ in the formula~(\ref{UnTE}) is equal to
\begin{equation}
E_{f}=\frac{\mu {u}^{2}}{2},  \label{UnTEf}
\end{equation}
where $\mu =m_{1} m_{2}/m_{0}$ is the reduced mass. It corresponds to the
energy in the center of mass frame required for ejecting fuel (fragment~1)
with the velocity $\mathbf{u}$ with respect to the rocket (fragment~2).
    Indeed, one can check that
    \begin{equation}
T_{1}^{(0)}+T_{2}^{(0)}=E_{f},  \label{vsT}
\end{equation}
where $T_{i}^{(0)}$ is the kinetic energy of particle ``i'' in the CM frame.
In the case of an ideal jet engine that converts all the energy of fuel into
that of jet stream, the value of $E_{f} $ would correspond just to this
energy of fuel stored in some form (chemical, nuclear, etc.).

    Let us denote the energy gain in the gravitational field
$E_{2\,\mathrm{grav}}$, and in the absence of gravitation $E_{2}^{(0)}$
(also for instantaneous ejection of fuel).
    Then, the difference between them is equal to
\begin{equation}
E_{2\,\mathrm{grav}}-E_{2}^{(0)}=\mu \mathbf{u}(\mathbf{v}_{b}-\mathbf{v}_{0}).
\label{g}
\end{equation}
Comparison under discussion implies that in both cases all masses and
velocities are the same in both cases (with and without gravitation), $E_0
\ge 0$ (so a rocket can start at infinity, $\mathbf{v}_0$ having the meaning of
the velocity there). Also, the initial states are chosen accordingly. Thus,
if we extend the tangent vector to the trajectory backward in time along the
straight line, it determines the initial position of a rocket in the
conditional scenario without gravitation. Then, the vector $\mathbf{v}_0$ is
the characteristic of a rocket in the scenario without gravitation, it is
pointed in the same direction as $\mathbf{v}_b$.
    The case when $\mathbf{u}\mathbf{v}_{b}>0$ corresponds to acceleration,
the case when $\mathbf{u}\mathbf{v}_{b}<0$ corresponds to deceleration.

    For collinear motion (when fuel is ejected along the trajectory in the same
direction),
\begin{equation}
E_{2\,\mathrm{grav}}-E_{2}^{(0)}=\mu u \left( \sqrt{v_{0}^{2} +
\frac{2GM}{r}} - v_{0} \right).  \label{uv}
\end{equation}
    Here and further we denote $ u = +|\mathbf{u}|$  % Hereafter
if $\mathbf{u}\mathbf{v}_{b} \ge 0$ and $ u = -|\mathbf{u}|$
if $\mathbf{u}\mathbf{v}_{b} < 0$.

    In the particular case, for $v_{0}=0$, the zero velocity
at infinity~(\ref{uv}) turns into
\begin{equation}
E_{2\,\mathrm{grav}}-E_{2}^{(0)}=\mu u\sqrt{\frac{2GM}{r}}=\mu uv_{p},
\label{E2nd0}
\end{equation}
where $v_{p}$ is the second cosmic (parabolic) velocity at distance~$r$ from
the attracting center.

    Now, let us estimate the efficiency of the gravitational Oberth effect
introducing the efficiency coefficient for the jet engine.
    There exist different definitions of this quantity in nonrelativistic
mechanics, see~\cite{AviatsiyaEnts}.
    We define the efficiency as the ratio of two
quantities. The first one is the increment of energy of a rocket without
account for ejected fuel.
    The second one is the sum of kinetic energy $ E_{k\,1}^{\,b} $
of fuel before turning engine on plus the stored thermal
(chemical, nuclear, etc.) energy of fuel~$E_{f}$:
\begin{equation}
\eta =\frac{|E_{2}-E_{2}^{b}|}{E_{1,k}^{\,b}+E_{f}}.  \label{nk14}
\end{equation}
Here, $E_{2}^{b}$ is the mechanical energy of a rocket (without account for
fuel) before turning on the engine, $E_{2}$ is that after turning it on. It
is worth noting that for an observer making measurements at rest in the
point where a rocket turns the engine on, the numerator of eq.~(\ref{nk14})
corresponds just to the absolute value of increment of energy, while the
denominator describes the consumed energy of fuel.
    Such a definition corresponds just to the so-called total efficiency of
jet engine~\cite{AviatsiyaEnts}.
    The modulus in the numerator of~(\ref{nk14}) guarantees
positive values of efficiency in the case of decelerating regime.

    Let us find efficiency for the process under consideration.
    Taking into account~(\ref{UnTE2}), (\ref{UnTEf}) and
\begin{equation}
E_{1,k}^{\,b}=\frac{m_{1}v_{b}^{2}}{2},\ \ \ \
E_{2}^{b}=\frac{m_{2}v_{b}^{2} }{2}-\frac{Gm_{2}M}{r},  \label{nk2}
\end{equation}
    we obtain
\begin{equation}
\eta =\frac{m_{2}}{m_{0}}\,|u|\frac{\left\vert \frac{m_{1}}{m_{0}}\frac{u}{2}
+\sqrt{\frac{2E_{0}}{m_{0}}+\frac{2GM}{r}}\right\vert }{\frac{m_{2}}{m_{0}}
\frac{u^{2}}{2}+\frac{E_{0}}{m_{0}}+\frac{GM}{r}}=\frac{m_{2}}{m_{0}}\,|u|\,
\frac{\left\vert \frac{m_{1}}{m_{0}}\frac{u}{2}+v_{b}\right\vert }{ \frac{
m_{2}}{m_{0}}\frac{u^{2}}{2}+\frac{v_{b}^{2}}{2}}.  \label{nk15}
\end{equation}
    For $v_{b}=0$, $\eta =m_{1}/m_{0}$ does not depend on the velocity of speed
with which fuel is being ejected.

    Considering the quantity~(\ref{nk15}) as a function $\eta (v_{b})$ of a
velocity $v_{b}$ of a rocket with fuel in the point where engine is turned
on, we find that for given $m_{1}$, $m_{2}$ the maximum is achieved
for $ v_{b}=um_{2}/m_{0}$.
    By substitution of this value into eq.~(\ref{nk15}), we
find after simple transformations that
\begin{equation}
\eta \left( \frac{m_{2}}{m_{0}}u\right) =\eta_{\,\mathrm{max}}=1.
\label{nk16}
\end{equation}
Then, taking into account~(\ref{v1b}) we can see that the velocity of
ejected fuel is equal to zero as well as its kinetic energy.
    According to~(\ref{nk14}), the total chemical energy $E_{f}$ stored
in fuel is spent to the increment of the kinetic energy of a rocket.

%%%%%%%%%%%%%%%%%%%%%%%%%%%%%%%%%%%%%%%%%%%%%%%%%%%%%%%%%%%%%%%%%%%%%

\section{Nonrelativistic analogue of the Penrose process}

    It follows from~(\ref{UnTE}) that
\begin{equation}
E_{2}-E_{0}-E_{f}=-E_{1} .
\end{equation}
    From (\ref{e0g})---(\ref{UnTEf}), one finds that
\begin{equation}
E_{1}=\frac{m_{1}}{m}E_{0}+\frac{m_{2}\mu }{2m}u^{2}-\mu \mathbf{v}_{b}\mathbf{u}.
\label{e1b}
\end{equation}
    In a similar way,
\begin{equation}
E_{2}=\frac{m_{2}}{m}E_{0}+\frac{m_{1}\mu }{2m}u^{2}+\mu \mathbf{v}_{b}\mathbf{u}.
\label{e2b}
\end{equation}

    Especially interesting case arises if $E_{1}<0$. Then, we gain more energy
than were invested, so we deal with the nonrelativistic analogue of the
Penrose process. Is it possible and under what conditions?

    Writing
\begin{equation}
\mathbf{v}_{b}\mathbf{u}=v_{b}\left\vert u\right\vert \cos \theta ,
\end{equation}
    equation~(\ref{e1b}) can be represented in the form
\begin{equation}
E_{1}=\frac{m_{1}}{m}E_{0}+\frac{m_{2}\mu }{2m}u^{2}-\mu v_{b}\left\vert
u\right\vert \cos \theta .  \label{e1u}
\end{equation}
Then, $E_{1}<0$, provided
\begin{equation}
u_{-}<\left\vert u\right\vert <u_{+},  \label{uu}
\end{equation}
\begin{equation}
u_{\pm }= \frac{m}{m_{2}} \left( v_{b}\cos \theta \pm \sqrt{\frac{2GM}{r}
-v_{b}^{2}\sin^{2}\theta } \right).
\end{equation}
This can be rewritten as
\begin{equation}
u_{\pm }=\frac{m}{m_{2}} \left( v_{b}\cos \theta \pm \sqrt{v_{b}^{2}
\cos^{2} \theta - v_{0}^{2}} \right).
\end{equation}
The expression under the square root should be nonnegative, so
\begin{equation}
\frac{2GM}{r}-v_{b}^{2}\sin^{2}\theta \geq 0,
\end{equation}
\begin{equation}
\sin^{2}\theta \leq \frac{v_{b}^{2}-v_{0}^{2}}{v_{b}^{2}} ,  \label{angle}
\end{equation}
    where we used~(\ref{vb}).
    Eq.~(\ref{uu}) implies that $u_{+}>0$, whence $ \cos \theta >0$.
    Therefore, the particle with $E_{1}<0$ can be ejected in
the hemisphere around the direction to motion of particle~0 only. In
particular, if $\theta =0$,
\begin{equation}
u_{\pm }=\frac{m}{m_{2}} \left( v_{b}\pm \sqrt{v_{b}^{2}-v_{0}^{2}} \right).
\end{equation}
For the circle orbit, eq.~(\ref{e0}) is not valid. In this case, as it is
known,
\begin{equation}
E_{0}=-\frac{GMm}{2r}, \ \ \ \ v_{b}^{2}=\frac{GM}{r} .
\end{equation}
Then, the condition $E_{1}<0$ entails
\begin{equation}
u_{\pm }=\frac{m}{m_{2}}\left( v_{b}\cos \theta \pm \sqrt{v_{b}^{2}\cos
^{2}\theta +\frac{GM}{r}}\right) =\frac{m}{m_{2}}\left( v_{b}\cos \theta \pm
\sqrt{v_{b}^{2}\cos^{2}\theta +v_{b}^{2}}\right) .
\end{equation}
    In this case, there is no restriction on the sign of $\cos \theta $, so the
particle with $E_{1}<0$ can be ejected in any direction.
    If we want to maximize $E_{2}$ (that is equivalent to minimizing $E_{1}$)
we must take $ \theta =0$.
    Then, all three vectors $\mathbf{v}_{b}$, $\mathbf{u}$, $\mathbf{v}_{1}$
and $\mathbf{v}_{2}$ are tangent to the trajectory.

    One can wonder how it is possible to speak about the Penrose process in
the nonrelativistic case as it is well known that Newtonian calculations can be
considered as a limit of the relativistic case where such a process is absent
for the Schwarzschild metric.
    However, there is no contradiction here.
    In this Section we discussed the properties of the nonrelativistic
energy~$E$ that contains the kinetic part and the potential energy.
    Meanwhile, the total energy contains also the rest mass.
    For the total energy there is no analogue of the Penrose process.

%%%%%%%%%%%%%%%%%%%%%%%%%%%%%%%%%%%%%%%%%%%%%%%%%%%%%%%%%%%%%%%%%%%%%%

\section{Relativistic case. General set-up}

    In the nonrelativistic case, we implied the conservation of mass according
to which $m_{0}=m_{1}+m_{2}$.
    Now, this condition cannot be fulfilled.
    Indeed, if we pass to the center of mass (CM) frame, we find that for any
nonzero relative velocity of fragments,
\begin{equation}
E_{0}=m_{0}c^{2}, \ \ \ E_{1}>m_{1}c^{2}, \ \ \ E_{2}>m_{2}c^{2}, \ \ \
E_{0}=E_{1}+E_{2}.
\end{equation}
    Therefore,
\begin{equation}
m_{0}>m_{1}+m_{2}.  \label{m12}
\end{equation}

    It is worth noting that the conservation law of energy~(\ref{UnTE}) in the
nonrelativistic case included also, additionally, chemical (or any other)
energy of jet fuel (or energy that is consumed in the decay of a particle).
In the relativistic case, this contribution enters automatically the total
energy of an original body $E_{0}$. In the nonrelativistic case the kinetic
energy and that of jet fuel were assumed to be small with respect to the
rest energies of objects. It is this circumstance that leads to the equality
$m_{0}=m_{1}+m_{2}$ in nonrelativistic approximation. Now, it is violated.

Now, we turn to consistent consideration of motion of a relativistic rocket
in the gravitational field. Let us consider particle motion in the
space-time with the metric
\begin{equation}
ds^{2}=-A(r)\,dt^{2}+\frac{dr^{2}}{A(r)}+r^{2}d\omega^{2},\ \ \ \ d\omega
^{2}=d\theta^{2}+\sin^{2}\theta \,d\varphi^{2}.  \label{g1}
\end{equation}
The energy of a particle with the mass $m$ in this metric can be found from
the formula (see eq.~(88.9) in~\cite{LL})
\begin{equation}
E=\frac{mc^{2}\sqrt{A}}{\sqrt{1-\frac{v^{2}}{c^{2}}}},  \label{en}
\end{equation}
    where $v$ is the particle velocity measured by a static observer with
fixed $ r,\theta ,\varphi $.
    In the static metric with the interval
$ ds^{2}=g_{00}dx_{0}^{2}+g_{\alpha \beta }dx^{\alpha }dx^{\beta }$,
where $ \alpha$, $\beta$ are spatial indices, we have
\begin{equation}
v=\frac{dl}{d\tau },\ \ \ \ d\tau =\frac{1}{c}\sqrt{g_{00}}\,dx^{0},\ \ \ \
dl^{2}=g_{\alpha \beta }\,dx^{\alpha }dx^{\beta }.
\end{equation}

Let a particle (rocket) moving freely in the gravitational field decay in
some point with the radial coordinate $r$ to two fragments (a rocket ejects
a portion of fuel).
    We use the same notations $E_{i},m_{i}$ ($i=0,1,2)$, $ u,v_{b} $
as in the nonrelativistic case.
    In particular, $u$ is a projection
of the velocity to the direction opposite to motion before decay.  We assume
that fuel is being ejected in the direction tangent to the trajectory (not
necessarily radial).  It is opposite to the direction of rocket motion in
the regime of acceleration~($u>0$) and is in the same direction in the
regime of deceleration~($u<0$).
    If $v_{b}=0$, we put $u = |u|$ and $v_{2} = |v_{2}|$.

The conservation law of energy and that of the projection of momentum to the
direction of motion in the point of ejection give us
\begin{equation}
E_{0}=E_{1}+E_{2},  \label{se1}
\end{equation}
    \begin{equation}
E_{0}v_{b}=E_{2}v_{2}+E_{1}v_{1}.  \label{se2}
\end{equation}
    According to the relativistic law of addition of velocities,
\begin{equation}
v_{1}=\frac{v_{2}-u}{1-\frac{\mathstrut v_{2}u}{c^{2}}}.  \label{se3}
\end{equation}

%%%%%%%%%%%%%%%%%%%%%%%%%%%%%%%%%%%%%%%%%%%%%%%%%%%%%%%%%%%%%%%%%%%%%%

\section{Scenario A, general formulas}

There are different scenarios depending on what data are assumed to be
fixed. In this scenario, we consider $E_{0}$, $m_{0}$, $m_{2}$ and $u$ as
known data. Then, the rest of quantities $E_{1}$, $E_{2}$, $v_{1}$, $v_{2}$
can be found from~(\ref{en}), (\ref{se1})--(\ref{se3}).

    For a particle with the energy $E$, mass $m$, moving in
the metric~(\ref{g1}), it follows from~(\ref{en}) that
    \begin{equation}
v=c\,\sqrt{1-\frac{A}{\varepsilon ^{2}}}\,,  \label{v}
\end{equation}
    where $\varepsilon =E/(mc^{2})$ is the specific energy,
\begin{equation}
Ev=c\,\sqrt{E^{2}-m^{2}c^{4}A(r)}\,.  \label{Ev}
\end{equation}
    After simple transformations, we obtain
\begin{equation}
\frac{v_{2}-u}{c\left( 1-\frac{v_{2}u}{c^{2}}\right) }= \frac{ \sigma _{v}
\sqrt{1-\frac{A}{\varepsilon _{2}^{2}}}\left( 1-\frac{u^{2}}{c^{2}}\right) -
\frac{A}{\varepsilon _{2}^{2}}\frac{u}{c}}{1-\frac{u^{2}}{c^{2}}\left( 1-
\frac{A}{\varepsilon _{2}^{2}}\right) }.  \label{v1o}
\end{equation}
    The sign factor $\sigma _{v}=+1$, if after turning on the engine, the
direction of rocket movement remains the same. If the fuel jet in the
deceleration regime was so big that the direction of rocker movement changed
to opposite, we put $\sigma _{v}=-1$.

    Eq.~(\ref{se2}) turns into
\begin{equation}
\frac{\varepsilon_{0}m_{0}}{\varepsilon_{2}m_{2}} \sqrt{1-
\frac{A}{ \varepsilon_{0}^{2}}} = \left( \frac{\varepsilon_{0}m_{0}}{
\varepsilon_{2}m_{2}}-1\right) \frac{\sigma_v \sqrt{1-\frac{A}{
\varepsilon_{2}^{2}}} \left( 1- \frac{u^{2}}{c^{2}}\right) -\frac{A}{
\varepsilon_{2}^{2}} \frac{u}{c}}{1-\frac{u^{2}}{c^{2}} \left( 1-\frac{A}{
\varepsilon_{2}^{2}}\right) } + \sigma_v \sqrt{1- \frac{A}{
\varepsilon_{2}^{2}}}.  \label{prsi}
\end{equation}
    We remind a reader that $u>0$ corresponds to the regime of acceleration
and $ u<0 $ corresponds to that of deceleration.

    For a photon rocket, substituting $u=\pm c$ in eq.~(\ref{prsi}), we obtain
\begin{equation}
\frac{\varepsilon_0 m_0 }{\varepsilon_2 m_2} \sqrt{ 1 - \frac{ A }{
\varepsilon_0^2} } = \sigma_u \left( 1 - \frac{\varepsilon_0 m_0 }{
\varepsilon_2 m_2} \right) + \sigma_v \sqrt{ 1 -
\frac{ A }{\varepsilon_2^2}} .  \label{fr}
\end{equation}
    Hereafter, $\sigma_u= +1$ refers to the regime of acceleration,
whereas $ \sigma_u= -1$ one corresponds to deceleration.
    The latter case implies that jet fuel (electromagnetic radiation)
is being ejected in the direction opposite to the motion of a rocket.

%%%%%%%%%%%%%%%%%%%%%%%%%%%%%%%%%%%%%%%%%%%%%%%%%%%%%%%%%%%%%%%%%%%%%%

\section{Photon rocket and efficiency in the relativistic case}

    Transforming~(\ref{fr}), one can obtain
\begin{equation}
\frac{\varepsilon_0 m_0 }{\varepsilon_2 m_2} \left( \sqrt{ 1 - \frac{ A }{
\varepsilon_0^2}} + \sigma_u \right) - \sigma_u = \sigma_v \sqrt{ 1 - \frac{
A }{\varepsilon_2^2} } .  \label{frp}
\end{equation}
    Squaring and simplifying, we find
\begin{equation}
E_2 = E_0 \left[ 1 - \frac{1}{2} \left( 1 - \frac{m_2^2}{m_0^2} \right)
\left( 1 - \sigma_u \sqrt{ 1 - A \left( \frac{ m_0 c^2 }{E_0} \right)^2 }
\right) \right] .  \label{E2fr}
\end{equation}
    whence
\begin{equation}
E_2 - E_0 = - \frac{E_0}{2} \left( 1 - \frac{m_2^2}{m_0^2} \right) \left( 1
- \sigma_u \sqrt{ 1 - A \left( \frac{ m_0 c^2 }{E_0} \right)^2 } \right).
\label{E2mE0Sch}
\end{equation}

    Now, we can estimate the efficiency of turning a photon rocket on in the
strong gravitational field, i.e. the Oberth effect. If such a rocket turns
the engine on near the horizon, $A\rightarrow 0$. Then, in the acceleration
regime, it follows from~(\ref{E2mE0Sch}) that $E_{2}-E_{0}\rightarrow 0$.
Thus the full energy of fuel transforms into the kinetic energy of a rocket.
    In this case, the efficiency tends to 100\thinspace \%.

When a particle approaches the horizon, its energy measured by a distant
observer tends to zero (infinite redshift). Thus the aforementioned equality
limiting value $E_{2}\rightarrow E_{0}$ complies with this property. And, it
is easy to show that a body radiates photons almost radially.

\begin{figure}[th]
\centering
\includegraphics[width=100mm]{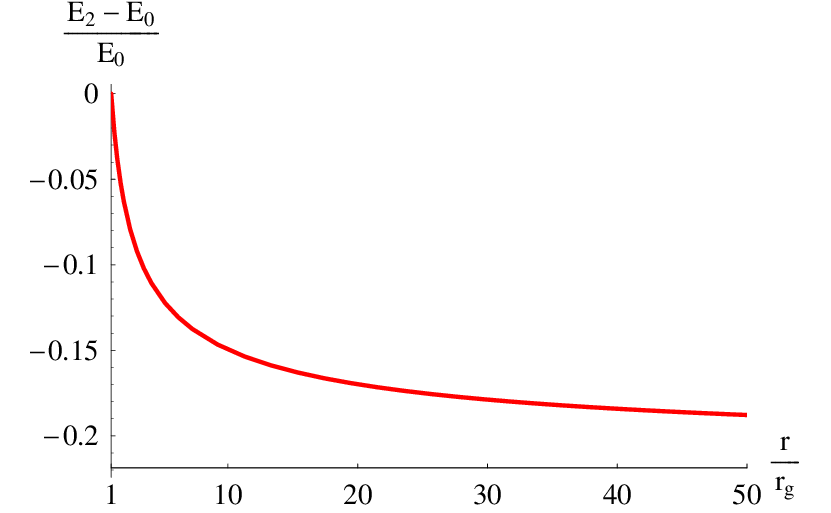}
\caption{ The plot $(E_2-E_0)/E_0$ of photon rocket in point with radial
coordinate~$r$ in the Schwarzschild metric for the case $E_0=m_0c^2$ and
$ m_2/m_0=0.75$.}
\label{FigGr1}
\end{figure}

The results of calculations for the case when decay of particle (turning the
engine on in the acceleration regime) occurs not on the horizon, are
presented on Fig.~\ref{FigGr1}.
%%%%%%%%%%%%%%%%%%%%%%%%%%%%%%%%%%%%%%%%%%%%%%%%%%
    In the absence of gravitational field ($A=1$)
\begin{equation}
E_{2}-E_{0}=\frac{(m_{2}^{2}-m_{0}^{2})c^{2}}{2m_{0}}\left( \frac{E_{0}}{
m_{0}c^{2}}-\sqrt{\left( \frac{E_{0}}{m_{0}c^{2}}\right)^{2} -1}\right) .
\label{E2mE0M}
\end{equation}
    For the case corresponding to Fig.~\ref{FigGr1},
this gives us $-7/16
\approx -0.22 $.

The role of the gravitational field is especially pronounced if $\Delta
m=m_{0}-m_{2}\ll m_{0}$.
    Thus, for $E_{0}=m_{0}c^{2}$ we find from~(\ref{E2mE0M})
\begin{equation}
E_{2}-E_{0}= - \Delta m c^{2} + \frac{(\Delta m)^{2}}{2m_{0}}c^{2}, \ \ \ \
\frac{E_{2}-m_{2}c^{2}}{\Delta mc^{2}}=\frac{\Delta m}{2m_{0}},
\label{E2delM}
\end{equation}
so only the fraction $(\Delta m)/(2m_{0})$ of the total energy of jet fuel
is used for the increase of the kinetic energy of a rocket. When $\Delta
m/m_{0}\rightarrow 0$, it vanishes. Meanwhile, in the same case near the
horizon the efficiency achieves 100\thinspace \%, as shown above.

%%%%%%%%%%%%%%%%%%%%%%%%%%%%%%%%%%%%%%%%%%%%%%%%%%%%%%%%%%

\section{Efficiency of relativistic rocket}

Now, we introduce the quantitative measure of efficiency $\eta $ for the jet
engine (in particular, a photon one) according to
\begin{equation}
\eta =\frac{|E_{2}-\varepsilon_{0}m_{2}c^{2}|}{\varepsilon_{0}(m_{0}-
m_{2})c^{2}}.  \label{kpd}
\end{equation}
    The quantity $\varepsilon_{0}m_{2}c^{2}$ gives us an initial energy of a
part of a rocket without fuel used due to turning on the engine.
    The factor $ \varepsilon_{0}$ takes into account that far from
an attracting body a rocket together with fuel moves with some initial
velocity (if $ \varepsilon_{0}>1$) or a rocket moves only within some bounded
region (if $ \varepsilon_{0}<1$).
    Then, the initial energy is equal to
\begin{equation}
E_{2}^{(0)}=\frac{m_{2}c^{2}}{\sqrt{1-\frac{v_{0}^{2}}{c^{2}}}}=
\varepsilon_{0}m_{2}c^{2} .
\end{equation}
    Therefore, the numerator equals the modulus of increment of the energy
without a used part of fuel. The quantity in the denominator of~(\ref{kpd})
describes the mass and the energy of fuel, where the factor $\varepsilon
_{0} $ takes into account the energy of motion. The denominator is the total
initial energy of used fuel. As in the relativistic case the energy of a
body includes the rest energy, the relativistic efficiency~$\eta $ does not
coincide with the quantity defined in~(\ref{nk14}).

    For the nonrelativistic case
\begin{equation}
E_{2}=\frac{m_{2}c^{2}\sqrt{g_{00}}}{\sqrt{1-\frac{v_{2}^{2}}{c^{2}}}}
\approx m_{2}\left( c^{2}+\frac{v_{2}^{2}}{2}-\frac{GM}{r}\right) ,\ \ \
\varepsilon_{0}\approx 1+\frac{v_{b}^{2}}{2c^{2}}-\frac{GM}{rc^{2}}
\label{EnR}
\end{equation}
taking into account~(\ref{v2b}) we get from~(\ref{kpd}) that
\begin{equation}
\eta =\frac{m_{2}}{m_{0}c^{2}}\left\vert \mathbf{v}_{b}\mathbf{u}+\frac{\Delta
mu^{2}}{2m_{0}}\right\vert =\frac{1}{m_{0}c^{2}}\left\vert m_{2}\mathbf{v}_{b}
\mathbf{u}+\frac{\mu u^{2}}{2}\right\vert .  \label{kpdnr}
\end{equation}
    Bearing in mind (\ref{UnTEf}), we can rewrite this formulas for the case of
acceleration ($\mathbf{v}_{b}\mathbf{u}>0$) as
\begin{equation}
\eta =\frac{E_{f}}{m_{0}c^{2}}+\frac{m_{2}\mathbf{v}_{b}\mathbf{u}}{m_{0}c^{2}}.
\label{ob}
\end{equation}
    The first term here represents the ratio of the energy stored in fuel (for
example, due to chemical or nuclear forces) to the initial energy at rest.
    The second term gives us correction just due to motion of a rocket.
    Thus, in the nonrelativistic case both terms are separated.
    If $\mathbf{v}_{b}$ and $ \mathbf{u}$ are parallel, this term is maximal
and is responsible for the Oberth effect which looks as an additive correction
to the first contribution.

For obtaining the value~(\ref{nk14}) in the nonrelativistic case, it is
necessary to subtract the rest energy of jet fuel in denominator
of~(\ref{kpd}).

    As in the Schwarzschild metric the states with negative energy are absent,
$ 0<E_{2}\leq E_{0}=\varepsilon_{0}m_{0}c^{2}$, the coefficient $\eta $
in~(\ref{kpd}) changes in limits $\eta \in \lbrack \,0,\,1]$.
    If $ E_{2}=E_{0}=\varepsilon_{0}m_{0}c^{2}$, it follows from~(\ref{kpd})
that $ \eta =1$ since all energy is used without loss. If the velocity of
a rocket without fuel is equal to that with fuel,
$\varepsilon_{0}=\varepsilon_{2}$.
    Then, $E_{2}=m_{2}\varepsilon_{2}=\varepsilon_{0}m_{2}$,
so $\eta =0$ in~(\ref{kpd}).

    From eq.~(\ref{E2fr}) for a photon engine and arbitrary $\varepsilon_{0}$
one obtains
\begin{equation}
\eta = \left| \frac{\Delta m}{2 m_0} + \sigma_u \left( 1 - \frac{\Delta m}{2
m_0} \right) \sqrt{ 1 - \frac{ A}{\varepsilon_0^2} } \right|.  \label{kpd2}
\end{equation}
    Let a rocket move in the acceleration regime ($\sigma_u= +1$). When it
approaches the attracting body, $A$ is decreasing, $\eta $ is increasing. In
the horizon limit $A\rightarrow 0$, $\eta \rightarrow 1$.

    It is worth noting that accounting for the rest mass of the source in the
nonrelativistic case would lead chemical sources of energy to the values of
efficiency of of the order $10^{-10}$ or less.
    For example, for the reaction
of combustion of hydrogen, specific heat of its combustion equals
141\,MJ/kg. This gives the value $1.6 \cdot 10^{-10}$ for the ratio of the
combusted energy to the rest energy of water that forms in this reaction. If
sources of energy are based on fission reaction of heavy nuclei, the ratio
of the released energy to the rest energy of initial radioactive substance
can reach $10^{-3} $. For light nuclei and the fusion reaction, the similar
ratio does not exceed $4 \cdot 10^{-3}$.

    At large distance from an attracting body, $\eta $ takes its minimum value
$ (\Delta m)/(2m_{0})$ for $E_{0}=m_{0}c^{2}$.

    For small $\Delta m$
\begin{equation}
\Delta m\rightarrow 0 \ \ \Rightarrow \ \ \eta =\sqrt{1-\frac{A}{
\varepsilon_{0}^{2}}}.  \label{mdm}
\end{equation}

    Now, we can estimate the value of the energy gain $E_{2\,\mathrm{grav}}$
for the acceleration of a photon rocket as compared to the similar
quantity $ E_{2}^{(0)}$ in the case of the absence of gravitation.
    As we are interested
now in the acceleration regime, we take the sign~$+$ in~(\ref{E2mE0Sch}).
In other words, we will estimate the value of the Oberth effect for a photon
rocket.
\begin{eqnarray}
E_{2\,\mathrm{grav}}-E_{2}^{(0)} &=&E_{0}\frac{(m_{0}^{2}-m_{2}^{2})}{
2m_{0}^{2}}\left( \sqrt{1-A\left( \frac{m_{0}c^{2}}{E_{0}}\right)^{2}}-
\sqrt{1-\left( \frac{m_{0}c^{2}}{E_{0}}\right)^{2}}\right) =  \notag \\
&=&\frac{1-A}{2m_{0}}\cdot \frac{(m_{0}^{2}-m_{2}^{2})c^{2}}{\sqrt{\left(
\frac{E_{0}}{m_{0}c^{2}}\right)^{2}-A}+\sqrt{\left( \frac{E_{0}}{m_{0}c^{2}}
\right)^{2}-1}}.  \label{E2fD0}
\end{eqnarray}
The relative gain for the photon engine equals
\begin{equation}
\frac{E_{2\,\mathrm{grav}}-E_{2}^{(0)}}{E_{2}^{(0)}}= \frac{
(m_{0}^{2}-m_{2}^{2})\left( \sqrt{1-A\left( \frac{m_{0}c^{2}}{E_{0}}
\right)^{2}}-\sqrt{1-\left( \frac{m_{0}c^{2}}{E_{0}}\right)^{2}}\right) }{
m_{0}^{2}+m_{2}^{2}+(m_{0}^{2}-m_{2}^{2})\sqrt{1-\left( \frac{m_{0}c^{2}}{
E_{0}}\right)^{2}}}.  \label{E2fDot}
\end{equation}

The formulas take especially simple form for the energy gain in the
Schwarzschild metric ($A=1-r_{g}/r$) and $E_{0}=m_{0}c^{2}$, as compared to
the instant ejection of photon fuel in the absence of gravitation $(A=1$)
\begin{equation}
E_{2\,\mathrm{grav}}-E_{2}^{(0)}=\frac{(m_{0}^{2}-m_{2}^{2})c^{2}}{2m_{0}}
\sqrt{\frac{r_{g}}{r}}.
\end{equation}
The relative gain, when $E_{0}=m_{0}c^{2}$ is equal to
\begin{equation}
\frac{E_{2\,\mathrm{grav}}-E_{2}^{(0)}}{E_{2}^{(0)}}=\frac{
m_{0}^{2}-m_{2}^{2}}{m_{0}^{2}+m_{2}^{2}}\sqrt{\frac{r_{g}}{r}}.
\label{E2frDot}
\end{equation}

%%%%%%%%%%%%%%%%%%%%%%%%%%%%%%%%%%%%%%%%%%%%%%%%%%%%%%%%%%%%%%%%%%%%%

\section{Rocket with nonphoton working medium}

In this case, eq.~(\ref{prsi}) with respect to $E_{2}$ can be written as an
algebraic equation of the sixth power. We will consider two particular cases
admitting simple approximate solutions.

%%%%%%%%%%%%%%%%%%%%%%%%%%%%%%%%%%%%%%%%%%%%%%%%%%

\subsection{\protect\normalsize Solution in the 1st order in $u/c$}

Neglecting terms of the order $u^{2}/c^{2}$, one obtains an approximate
solution of~(\ref{prsi}):
\begin{equation}
E_{2} \approx E_{0}\frac{m_{2}}{m_{0}}\left[ 1+\frac{u}{c} \left( 1-\frac{
m_{2}}{m_{0}}\right) \sqrt{1-A\left( \frac{m_{0}c^{2}}{E_{0}} \right)^{2}}
\right].  \label{E21u}
\end{equation}
In this case, the efficiency
\begin{equation*}
\eta =\frac{|u|}{c}\frac{m_{2}}{m_{0}}
\sqrt{ 1- \frac{A}{\varepsilon_{0}^{2}}} .
\end{equation*}
The energy gain in the gravitational field in the acceleration regime, as
compared to that for fuel ejection in the absence of gravitation is equal to
    \begin{equation}
E_{2\,\mathrm{grav}}-E_{2}^{(0)} \approx \frac{m_{2}(m_{0}-m_{2}) }{m_{0}}
cu \left( \sqrt{\varepsilon_{0}^{2}-A}-\sqrt{\varepsilon_{0}^{2}-1} \right).
\label{E21uD}
\end{equation}
    When $E_{0}=m_{0}c^{2}$, eq.~(\ref{E2nd0}) is reproduced.

%%%%%%%%%%%%%%%%%%%%%%%%%%%%%%%%%%%%%%%%%%%%%%%%%%%%%%%%%%%%%%%%%%%%%%

\section{Solution near horizon}

\begin{figure}[h]
\centering
\includegraphics[width=80mm]{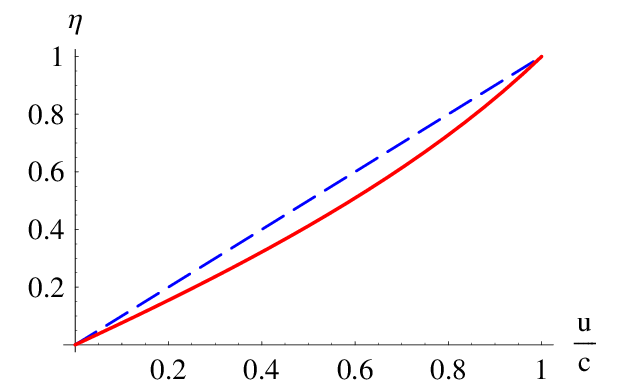}
\caption{The plot of efficiency near event horizon in dependence of reactive
fuel velocity for $m_2 \approx m_0$ (the blue dashed line) and $m_2 = 0.75
m_0$ (the red line).}
\label{Figkpd}
\end{figure}

For an arbitrary value of $u/c$, the solution of eq.~(\ref{prsi}) in the
first approximation in $A$ takes the form
\begin{equation}
E_{2}{\approx }E_{0}\frac{1+\frac{u}{c}}{\frac{u}{c}+\sqrt{\frac{u^{2}}{
c^{2} }+\frac{m_{0}^{2}}{m_{2}^{2}}\left( 1-\frac{u^{2}}{c^{2}}\right) }},
\label{EA}
\end{equation}
\begin{equation}
E_{2}-E_{0} \approx -E_{0}\frac{\left( 1-\frac{u}{c}\right) \left( \frac{
m_{0}^{2}}{m_{2}^{2}}-1\right) }{\frac{u}{c}+ \sqrt{\frac{u^{2}}{ c^{2}}+
\frac{m_{0}^{2}}{m_{2}^{2}} \left( 1-\frac{u^{2}}{c^{2}}\right) } + \frac{
m_{0}^{2}}{m_{2}^{2}} \left( 1-\frac{u}{c}\right) }.  \label{EAd}
\end{equation}
The efficiency~(\ref{kpd}) is equal now
\begin{equation}
\eta =\frac{u}{c}\frac{\frac{u}{c}+\frac{m_{0}}{m_{2}}\left( 1+\frac{u}{c}
\right) +\sqrt{\frac{u^{2}}{c^{2}}+\frac{m_{0}^{2}}{m_{2}^{2}}\left( 1-\frac{
u^{2}}{c^{2}}\right) }}{\left( \frac{m_{0}}{m_{2}}+\sqrt{\frac{u^{2}}{c^{2}}
+ \frac{m_{0}^{2}}{m_{2}^{2}}\left( 1-\frac{u^{2}}{c^{2}}\right) }\right)
\left( \frac{u}{c}+\sqrt{\frac{u^{2}}{c^{2}}+\frac{m_{0}^{2}}{m_{2}^{2}}
\left( 1-\frac{u^{2}}{c^{2}}\right) }\right) }.  \label{kpdA}
\end{equation}
When $u\rightarrow 0$ the efficiency also tends to zero. For a photon
rocket, $u/c=1$, the efficiency $\eta \rightarrow 1$ near the horizon that
agrees with the results described above. If the quantity of ejected fuel is
negligible, so $m_{2}\approx m_{0}$, we obtain from~(\ref{kpdA}) that $\eta
=|u|/c $. The case when the fraction of ejected fuel is substantial, is
plotted on Fig.~\ref{Figkpd}.
%%%%%%%%%%%%%%%%%%%%%%%%%%%%%%%%%%%%%%%%%%%%%%%%%%

%%%%%%%%%%%%%%%%%%%%%%%%%%%%%%%%%%%%%%%%%%%%%%%%%%%%%%%%%%%%%%%%%%%%%%

\section{General solution for small $\Delta m$}

Now, let us find a general solution of eq.~(\ref{prsi}) for a nonphoton
rocket for the case $\Delta m/m_{0}\ll 1$. We denote
\begin{equation}
\frac{m_{2}}{m_{0}}=1-\alpha ,\ \ \ \alpha \ll 1,  \label{nk17}
\end{equation}
\begin{equation}
\frac{\varepsilon_{2}}{\varepsilon_{0}}=1+\beta .  \label{nk18}
\end{equation}
It is worth noting that the efficiency~(\ref{kpd}) in this case equals
\begin{equation}
\eta =\frac{|\beta |}{\alpha }.  \label{nk19}
\end{equation}

We will look for $\beta $ as a function of $\alpha $. Assuming that for
small $\alpha $, the quantity $|\beta |\ll 1$, we carry out calculations in
the first order with respect to~$\alpha $ and~$\beta $:
\begin{equation}
\frac{\varepsilon_{0}m_{0}}{\varepsilon_{2}m_{2}}=1+\alpha -\beta ,
\label{nk20}
\end{equation}
\begin{equation}
\frac{A}{\varepsilon_{2}^{2}}=\frac{A}{\varepsilon_{0}^{2}}(1-2\beta ),\ \ \
\ 1-\frac{A}{\varepsilon_{2}^{2}}=1-\frac{A}{\varepsilon_{0}^{2}}+2\beta
\frac{A}{\varepsilon_{0}^{2}},  \label{nk22}
\end{equation}
\begin{equation}
\sqrt{1-\frac{A}{\varepsilon_{2}^{2}}} = \left[
\begin{array}{cl}
\displaystyle \sqrt{\mathstrut 2\beta }\ ,\  & \displaystyle \frac{A}{%
\varepsilon_{0}^{2}}=1, \\[11pt]
\displaystyle \sqrt{1-\frac{A}{\varepsilon_{0}^{2}}}+\frac{\beta A}{
\varepsilon_{0}^{2} \sqrt{1-\frac{A}{\varepsilon_{0}^{2}}}}\ ,\  & %
\displaystyle \frac{A}{\varepsilon_{0}^{2} } \neq 1.%
\end{array}
\right.  \label{nk23}
\end{equation}
If $A/\varepsilon_{0}^{2}=1$, a rocket remains at rest before turning the
engine on, hence the deceleration regime in this case is impossible, $\beta
\geq 0$. Then, in the main approximation eq.~(\ref{prsi}) takes the form
\begin{equation}
0=-(\alpha -\beta )\frac{u}{c}+\sqrt{2\beta }.  \label{nk24}
\end{equation}
As a result, we obtain
\begin{equation}
\frac{A}{\varepsilon_{0}^{2}}=1\ \ \Rightarrow \ \ \beta =\alpha^{2}\frac{%
u^{2}}{2c^{2}},\ \ \eta =\alpha \frac{u^{2}}{2c^{2}}, \ \ \ \alpha
\rightarrow 0.  \label{nk25}
\end{equation}

If $A/\varepsilon_{0}^{2}<1$, it follows from~(\ref{prsi}) that
\begin{equation}
\frac{A}{\varepsilon_{0}^{2}}<1\ \ \Rightarrow \ \ \beta =\alpha \frac{u}{c}
\sqrt{1-\frac{A}{\varepsilon_{0}^{2}}},\ \ \eta =\frac{|u|}{c}\sqrt{1-
\frac{A}{\varepsilon_{0}^{2}}},\ \ \ \alpha \rightarrow 0.  \label{nk26}
\end{equation}
On the horizon, it follows from~(\ref{nk26}) that the efficiency equals
\begin{equation}
A=0,\ \ \ \frac{\Delta m}{m_{0}}\ll 1\ \ \Rightarrow \ \ \eta =\frac{|u|}{c}.
\label{nk27}
\end{equation}

%%%%%%%%%%%%%%%%%%%%%%%%%%%%%%%%%%%%%%%%%%%%%%%%%%%%%%%%%%%%%%%%%%%%%%

\section{Scenario B}

    Now, we can somewhat change the set of quantities which we assumed to be
fixed. Namely, let us fix the velocity of particle~1 (fuel)
$\mathbf{v}_{1}^{(0)}$ in the CM frame instead of
the relative velocity~$\mathbf{u}$.
    We also fix the mass $m_{1}$.
    Hereafter, we call it \textquotedblleft scenario~B\textquotedblright .
    For the nonrelativistic case, the relation
between scenarios~A and B is quite direct. To relate them, one should
express $\mathbf{v}_{2}^{(0)}$, $\mathbf{v}_{1}$, $\mathbf{v}_{2}$ and $u$
in terms of $\mathbf{v}_{1}^{(0)}$ and and put $m_{2}=m_0-m_{1}$.
    As this procedure is
quite direct, we omit corresponding simple formulas. However, for the
relativistic case, the situation becomes much more interesting, with
relation between scenarios~A and B being nontrivial. It is especially
important that scenario~B (in contrast to scenario~A) admits simple exact
algebraic solutions for all relevant quantities (see below).

Hereafter, we omit the factor $c$ if this does not lead to confusion. Let us
again consider the decay of particle~0. When new particles~1 and 2 move
under some angles to the original trajectory, this leads to very cumbersome
formulas in relations between the center of mass (CM) frame and the static
one. To simplify matter, we assume that (i) decay of particle 0 occurs just
in the turning point, (ii) new particles~1 and 2 fly out along the tangent
direction to the same trajectory. To avoid confusion because of using
superscript ``0'', hereafter we denote components of the velocity in the CM
frame $\tilde{v}_{1}$, $\tilde{v}_{2}$.

    Then, the standard relativistic formulas of additions of velocities give us
\begin{equation}
v_{2}=\frac{v_{b}+\tilde{v}_{2}}{1+v_{b}\tilde{v}_{2}} ,  \label{v2v}
\end{equation}
\begin{equation}
v_{1}=\frac{v_{b}+\tilde{v}_{1}}{1-v_{b}\left\vert \tilde{v}_{1}\right\vert }
.
\end{equation}
By definition, the momentum in the CM frame vanishes, the total energy is
equal to the energy in the CM frame $m_{0}$, so we have two equations
\begin{equation}
\frac{m_{1}\tilde{v}_{1}}{\sqrt{1-\tilde{v}_{1}^{2}}}+\frac{m_{2}\tilde{v}
_{2}}{\sqrt{1-\tilde{v}_{2}^{2}}}=0 ,
\end{equation}
\begin{equation}
\frac{m_{1}}{\sqrt{1-\tilde{v}_{1}^{2}}}+\frac{m_{2}}{\sqrt{1-
\tilde{v}_{2}^{2}}}=m_{0} .
\end{equation}
For definiteness, we assume that $\tilde{v}_{2}>0$, $\tilde{v}_{1}<0$ that
corresponds to the regime of acceleration of a rocket (which coincides with
particle 0 before ejection of fuel and with particle 2 after it).

It follows form these equations that
\begin{equation}
\tilde{E}_{2}=m_{0}-\tilde{E}_{1} ,
\end{equation}
\begin{equation}
m_{2}=\sqrt{m_{0}^{2}+m_{1}^{2}-2m_{0}\tilde{E}_{1}},  \label{m2p}
\end{equation}
\begin{equation}
\tilde{v}_{2}=-\frac{\tilde{P}_{1}}{m_{0}-\tilde{E}_{1}} ,  \label{v2p}
\end{equation}
where
\begin{equation}
\tilde{P}_{1}=\frac{m_{1}\tilde{v}_{1}}{\sqrt{1-\tilde{v}_{1}^{2}}}
=-\left\vert \tilde{P}_{1}\right\vert ,\ \ \ \tilde{E}_{1}=\frac{m_{1}}{
\sqrt{1-\tilde{v}_{1}^{2}}}.
\end{equation}

Usually, in textbooks $\tilde{E}_{1}$ and $\tilde{E}_{2}~$are expressed in
terms of masses (see e.g. Sec.~II,~11 in~\cite{LL}).
    But now we consider $ \tilde{E}_{1}$ and $m_{1}$ as given and find
characteristics of particle~2.
    Then, (\ref{v2v}) gives us
\begin{equation}
v_{2}=\frac{v_{b}+\frac{\left\vert \tilde{P}_{1}\right\vert }{m_{0}-
\tilde{E}_{1}}}{1+v_{b}\frac{\left\vert \tilde{P}_{1}\right\vert }{m_{0}-
\tilde{E}_{1} }} .
\end{equation}

In the static frame, we have from~(\ref{en}) and~(\ref{v2v})
\begin{equation}
E_{2}=\frac{\tilde{E}_{2}+v_{b}\left\vert \tilde{P}_{2}\right\vert }{ \sqrt{
1-v_{b}^{2}}}\sqrt{A}
\end{equation}
that generalizes the Lorentz formula typical of the flat space-time. It can
be also written as
\begin{equation}
E_{2}=\frac{m_{0}-\tilde{E}_{1}+v_{b}\left\vert \tilde{P}_{1}\right\vert }{
\sqrt{1-v_{b}^{2}}}\sqrt{A}.
\end{equation}
In a similar way,
\begin{equation}
E_{1}=\frac{\tilde{E}_{1}-v_{b}\left\vert \tilde{P}_{1}\right\vert }{
\sqrt{1-v_{b}^{2}}}\sqrt{A}.  \label{113}
\end{equation}
Obviously, $E_{1}+E_{2}=E_{0},$ as it should be. We also find
\begin{equation}
\varepsilon_{2}\equiv \frac{E_{2}}{m_{2}}=\frac{m_{0}-\tilde{E}
_{1}+v_{b}\left\vert \tilde{P}_{1}\right\vert }{\sqrt{1-v_{b}^{2}}
\sqrt{m_{0}^{2}+m_{1}^{2}-2m_{0}\tilde{E}_{1}}}\sqrt{A}.  \label{114}
\end{equation}
Here,
\begin{equation}
\tilde{E}_{1}\leq \frac{m_{0}^{2}+m_{1}^{2}}{2m_{0}}.  \label{Emm}
\end{equation}

%%%%%%%%%%%%%%%%%%%%%%%%%%%%%%%%%%%%%%%%%%%%%%%%%%%%%%%%%%%%%%%%%%%%%%

\section{Photon rocket}

    We can take the safe limit to the case of photon fuel, provided
$ m_{1}\rightarrow 0$, $\tilde{v}_{1}\rightarrow 1$ in such a way that
$\tilde{E}_{1}$ remains finite, $\tilde{P}_{1}=\tilde{E}_{1}$.
    Then,
\begin{equation}
\tilde{v}_{2}=\frac{\tilde{E}_{1}}{m_{0}-\tilde{E}_{1}} .
\end{equation}
The condition $\left\vert \tilde{v}_{2}\right\vert <1$ gives us
\begin{equation}
\tilde{E}_{1}\leq \frac{m_{0}}{2}  \label{ineq}
\end{equation}
in agreement with~(\ref{Emm}). We also find from~(\ref{m2p}) that
\begin{equation}
m_{2}=\sqrt{m_{0}^{2}-2m_{0}\tilde{E}_{1}} .  \label{m2}
\end{equation}
For particle~1 we have from~(\ref{113})
\begin{equation}
E_{1}=\tilde{E}_{1}\sqrt{A}\sqrt{\frac{1-v_{b}}{1+v_{b}}}.
\end{equation}
    This is just combination of the Doppler shift and redshift ---
see~\cite{zero} for details and references therein.
    It follows from~(\ref{114}), (\ref{m2}) that
\begin{equation}
E_2 = E_0 - \tilde{E}_{1}\sqrt{A}\sqrt{\frac{1-v_{b}}{1+v_{b}}}.  \label{120}
\end{equation}

%%%%%%%%%%%%%%%%%%%%%%%%%%%%%%%%%%%%%%%%%%%%%%%%%%%%%%%%%%%%%%%%%%%%%%

\subsection{Efficiency}

Let us use the definition of the efficiency according to eq.~(\ref{kpd}).
Then it follows from~(\ref{en}), (\ref{se1}) and~(\ref{m2}) that
\begin{equation}
\eta =\frac{E_{0}-\tilde{E}_{1} \sqrt{A}\sqrt{\frac{1-\sqrt{1-\frac{A}{
\varepsilon_{0}^{2}}}}{1+\sqrt{1-\frac{A}{\varepsilon_{0}^{2}}}}}
-\varepsilon_{0}\sqrt{m_{0}^{2}-2m_{0}\tilde{E}_{1}}}{\varepsilon_{0} \left(
m_{0}-\sqrt{m_{0}^{2}-2m_{0}\tilde{E}_{1}} \right)} .
\end{equation}
In the horizon limit $A\rightarrow 0$ we have
\begin{equation}
\eta =1 .
\end{equation}

    Let us consider the nonrelativistic case $\varepsilon_{0}=1$,
$ v_{b}\rightarrow 0$, $A=1$ (no gravitation).
    Then,
\begin{equation}
\eta =\frac{E_{0}-\tilde{E}_{1}-\varepsilon_{0}m_{2}}{ \left( m_{0} -
\sqrt{ m_{0}^{2}-2m_{0}\tilde{E}_{1}} \right)}.
\end{equation}
    One can check that~(\ref{120}) agrees with~(\ref{E2fr}) for a given choice
of signs (acceleration regime), if one take into account eq.~(\ref{m2p}).
Thus the efficiency found in Scenario B agrees with that analyzed in Sec.~7
within Scenario~A.

%%%%%%%%%%%%%%%%%%%%%%%%%%%%%%%%%%%%%%%%%%%%%%%%%%%%%%%%%%%%%%%%%%%%%%

\section{Non-collinear vs collinear motion}

    Up to now, we considered the case of collinear motion, when both particles
1, 2 are emitted in the tangent direction of the trajectory of particle~0.
Now, we consider a more general case, when all three particles have
arbitrary nonzero angular momenta $L_{i}$ ($i=0,1,2$) with the restriction
$ L_{0}=L_{1}+L_{2}$.

    According to eqs.~(19) --- (30) of~\cite{centr},
\begin{equation}
E_{1}=\frac{1}{2\tilde{m}_{0}^{2}} \left( E_{0}\Delta_{+}+P_{0}\delta
\sqrt{D} \right),  \label{e1}
\end{equation}
\begin{equation}
E_{2}=\frac{1}{2\tilde{m}_{0}^{2}} \left(E_{0}\Delta_{-}-P_{0}\delta \sqrt{D}
\right),  \label{e2}
\end{equation}
$\delta =\pm 1$. Here, we use particle labels 1 and 2 instead of 3 and 4
respectively in~\cite{centr}. In this section, we put $c=1$ for simplicity.
    \begin{equation}
\Delta_{\pm }=\tilde{m}_{0}^{2}\pm (\tilde{m}_{1}^{2}-\tilde{m}_{2}^{2}),
\label{del}
\end{equation}
    \begin{equation}
\tilde{m}_{i}^{2}=m_{i}^{2}+\frac{L_{i}^{2}}{r^{2}} ,  \label{mm}
\end{equation}
where $i=0,1,2$,
    \begin{equation}
D=\Delta_{+}^{2}-4\tilde{m}_{0}^{2}\tilde{m}_{1}^{2}=\Delta_{-}^{2}-
4 \tilde{m}_{0}^{2}\tilde{m}_{2}^{2} .  \label{D}
\end{equation}

    The direction of motion is characterized by a quantity $\sigma $,
where $ \sigma =+1$ for motion in the outward direction and $\sigma =-1$
for the inward case.
    We are mainly interested in the situation, when particle~0 (an initial
rocket) moves towards a black hole, from large radii to smaller ones.
    Then, at least one of particles falls in a black hole.
    We assume that this is particle 2, so $\sigma_{2}=-1$.

    As is explained in~\cite{centr}, possible scenarios can be characterized by
the set $(\sigma_{1}$, $e_{1}$, $e_{2}$, $\delta $).
\begin{equation}
e_{i}=sgn \left( \Delta_{+}\sqrt{A}-2\tilde{m}_{i}E_{0} \right) ,
\end{equation}
    where $i=1,2$. These quantities were denoted $\varepsilon_{i}$
in~\cite{centr}.

Then, there exist 6 scenarios. They are listed in eq.~(30) of~\cite{centr}.
We are interested in the process close to the horizon, where $A$ is
sufficiently small. Let us assume that $\tilde{m}_{i}\neq 0$
(either $ m_{i}\neq 0$ or $L_{i}\neq 0$ or both).
    Then, $e_{1}=e_{2}=-1$ and only the
following scenarios survive. If $\delta =+1,$ only scenario $(-,-,-,+)$ is
possible. If $\delta =-1$, we have $(-,-,-,-)$.
    Now, $ \sigma_{1}= \sigma_{2}=-1$, both particles fall in a black hole.
    Two scenarios differ only by labels~1 and~2.

%%%%%%%%%%%%%%%%%%%%%%%%%%%%%%%%%%%%%%%%%%%%%%%%%%

\subsection{The most efficient configuration}

We are interested in the scenario that gives us the maximum possible value
of $E_{2}$ for given other data. It follows from intuitively clear arguments
that for the decay to be the most efficient, particle 1 must be ejected in
the direction strictly opposite to the direction in which an initial
particle 0 moves. For the nonrelativistic case, this can be justified very
easily - see Sec. \ref{turn}. However, in the relativistic case, instead of
the nonrelativistic one, the situation is less obvious because of nonlinear
character of relevant quantities. Below, we give an explicit proof of the
aforementioned statement.

    Let all masses $m_{i}$, $E_{0}$ and the angular momentum $L_{0}$ be fixed.
    We have at our disposal one independent variable, this is the momentum
$ L_{2} $ (then $L_{1}=L_{0}-L_{2}$).
    The condition of the extremum of $E_{2}$ entails
\begin{equation}
\frac{\partial E_{2}}{\partial L_{2}}=0 .  \label{de}
\end{equation}
    Then, using (\ref{e1})---(\ref{D}), one can obtain from~(\ref{de}) the
equation
\begin{equation}
E_{0}L_{0}\sqrt{D}=P_{0}\delta \left( L_{0}b-2L_{2}m_{0}^{2} \right) .
\label{eld}
\end{equation}

    On the other hand, we can consider the scenario under discussion from the
geometric viewpoint.
    If we want motion of debris to be parallel to the
initial particle, the corresponding angles should coincide. Assuming all
particles to move in the same plane, we see that the ratio $v^{(3)}/v^{(1)}$
of tetrad components of the three-velocity corresponding to a static
observer should have the same value for particles~0 and~2.
    Here, indices~1 and~3 tetrads pointed in the radial and angle directions,
respectively.
    It is easy to find that
\begin{equation}
v^{(3)}=\frac{L\sqrt{A}}{rE} ,
\end{equation}
\begin{equation}
v^{(1)}=\sqrt{1-\frac{A\tilde{m}^{2}}{E^{2}}}.
\end{equation}
This can be derived from the equations of geodesic motion or taken directly
from eqs.~(12), (13) of~\cite{k}. Then,
\begin{equation}
\frac{L_{0}}{P_{0}}=\frac{L_{2}}{\left\vert P_{2}\right\vert } ,  \label{LP}
\end{equation}
$\left\vert P_{2}\right\vert =\sqrt{E_{2}^{2}-\tilde{m}_{2}A}$.
    Using~(\ref{e2}), one can find that
\begin{equation}
\left\vert P_{2}\right\vert =\frac{E_{0}\sqrt{D}+P_{0}\Delta_{-}\delta }{
2 \tilde{m}_{0}^{2}}
\end{equation}
    in agreement with eqs.~(26), (27) of~\cite{centr}.
    Then, after some algebra, it is easy to show that~(\ref{LP}) is equivalent
to~(\ref{eld}). Thus the statement about the most efficient scenario is proven.

    Now, we can solve eq.~(\ref{eld}).
    After some algebraic manipulations, one finds
\begin{equation}
L_{2}=\frac{L_{0}}{2m_{0}^{2}} \left( b\pm \frac{E_{0}}{\sqrt{E_{0}^{2}-
m_{0}^{2}A}} \sqrt{b^{2}-4m_{0}^{2}m_{2}^{2}} \right), \ \ \ \
b=m_{0}^{2}+m_{2}^{2}-m_{1}^{2} ,
\end{equation}
\begin{equation}
E_{2}=\frac{E_{0}b}{2m_{0}^{2}}\pm \frac{\sqrt{E_{0}^{2}-m_{0}^{2}A}}{
2m_{0}^{2}}\sqrt{b^{2}-4m_{0}^{2}m_{2}^{2}} .  \label{E2L}
\end{equation}
    Here, the upper sign refers to the acceleration regime, the lower one -- to
the deceleration regime.

    If $m_{1}=0$ (a photon rocket), eq.~(\ref{E2L}) coincides
with eq.~(\ref{E2fr}).

%%%%%%%%%%%%%%%%%%%%%%%%%%%%%%%%%%%%%%%%%%%%%%%%%%

\subsection{Near-horizon expansion}

Near the horizon when $A\rightarrow 0$,
\begin{equation}
P_{0}=E_{0}-\frac{A\tilde{m}_{0}^{2}}{2E_{0}}+O(A^{2}),
\end{equation}
\begin{equation}
L_{2} \approx \frac{ L_{0} }{2 m_0^2} \left( b\pm \sqrt{ b^2 - 4 m_0^2
m_2^{2}} \right),  \label{n112}
\end{equation}
\begin{equation}
E_{2}\approx \frac{E_{0}}{2m_{0}^{2}} \left( b\pm \sqrt{b^{2}-4m_{0}^{2}
m_{2}^{2}} \right) .  \label{n113}
\end{equation}
According to~(\ref{kpd}), in this limit the efficiency in the acceleration
regime
\begin{equation}
\eta =\frac{m_{0}}{m_{0}-m_{2}}\left( \frac{b+\sqrt{b^{2}-4m_{0}^{2}m_{2}^{2}
}}{2m_{0}^{2}}-\frac{m_{2}}{m_{0}}\right) .
\end{equation}
    For a photon rocket $m_{1}=0$ and $\eta \rightarrow 1$ in agreement
with~(\ref{kpd2}).

%%%%%%%%%%%%%%%%%%%%%%%%%%%%%%%%%%%%%%%%%%%%%%%%%%%%%%%%%%%%%%%%%%%%%%

\section{Decay in the turning point\label{turn}}

Let us consider the decay of particle~0. When new particles~1 and~2 move
under some angles to the original trajectory, this leads to very cumbersome
formulas. To simplify matter and concentrate on the physically relevant
situation, we discuss now the following scenario. (i)~Decay of particle~0
occurs just in the turning point, (ii)~new particles~1 and~2 fly out along
the tangent direction to the same trajectory. This means that the radial
velocity of each particle vanishes, so $P_{0}=P_{1}=P_{2}=0$. Then,
according to eqs.~(\ref{e1})--(\ref{D}) in the point of decay $r_{d}$,
\begin{equation}
E_{i}=\tilde{m}_{i}\sqrt{A_{d}}=m_{i}\sqrt{A_{d}}\sqrt{1+\frac{x_{i}^{2}}{
m_{i}^{2}}} ,  \label{ea}
\end{equation}
where $x_{i}\equiv L_{i} / r_{d}$, $i=0,1,2$, $A_{d}=A(r_{d})$, where we
used the definition~(\ref{mm}). From the conservation of energy, we obtain
\begin{equation}
\tilde{m}_{0}=\tilde{m}_{1}+\tilde{m}_{2} .  \label{118}
\end{equation}
It follows from~(\ref{118})
\begin{equation}
x_{1}=\frac{d_{+}}{2m_{0}^{2}}x_{0}\mp \frac{\tilde{m}_{0}}{2m_{0}^2}
\sqrt{ D_{0}} ,
\end{equation}
\begin{equation}
x_{2}=\frac{d_{-}}{2m_{0}^{2}}x_{0}\pm \frac{\tilde{m}_{0}}{2m_{0}^2}
\sqrt{ D_{0}} ,  \label{x2}
\end{equation}
\begin{equation}
D_{0}=d_{+}^{2}-4m_{1}^{2}m_{0}^{2}=d_{-}^{2}-4m_{2}^{2}m_{0}^{2} ,
\end{equation}
\begin{equation}
d_{+}=m_{0}^{2}+m_{1}^{2}-m_{2}^{2}, \ \ \ \
d_{-}=m_{0}^{2}+m_{2}^{2}-m_{1}^{2}.
\end{equation}
    These formulas resemble those for the energy in Sec. IV of~\cite{centr},
with $\tilde{m}_{i}$ replaced with $m_{i}$ in proper places.

    A velocity of particle $i$ is directed along the $\phi $ axis. Using the
tetrads attached to a static observer, one easily finds that in the turning
point. We obtain that (e.g., see eq.~(44) in Ref.~\cite{TZ21}) its value
    \begin{equation}
v_{i}^{(3)}=\frac{\sqrt{A}\mathcal{L}_{i}}{r\varepsilon_{i}}=
\frac{\sqrt{ E_{i}^{\mathstrut 2}-m_{i}^{2}A}}{E_{i}}=\sqrt{1-
\frac{m_{i}^{2}}{ \tilde{m}_{i}^{2}}} ,
\end{equation}
    where $ \mathcal{L}_{i} = L_i/m_i$.

In the acceleration regime we have for particle~2 (rocket) we should take
the upper sign in~(\ref{x2}). Below, we assume in this section that a rocket
is photonic, so $m_{1}=0$. Then, we have from~(\ref{x2}) in the acceleration
regime (assuming $x_{0}>0$)
\begin{equation}
x_{1}=\frac{x_{0}}{2}(1-\alpha^{2})-\frac{m_{0}}{2}\sqrt{1+\frac{x_{0}^{2}}{
m_{0}^{2}}}(1-\alpha^{2})\equiv m_{0}g_{-}(\frac{x_{0}}{m_{0}},\alpha ).
\label{f1}
\end{equation}
\begin{equation}
\frac{x_{2}}{m_{2}}=\frac{x_{0}}{2m_{0}}\frac{1}{\alpha }(1+\alpha^{2})+
\frac{1}{2\alpha }\sqrt{1+\frac{x_{0}^{2}}{m_{0}^{2}}}(1-\alpha^{2})\equiv
f_{+}(\frac{x_{0}}{m_{0}} , \alpha ),  \label{f2}
\end{equation}
$\alpha \equiv m_2 /m_0 $. By substitution into~(\ref{ea}), we obtain
\begin{equation}
\varepsilon_{2}=\sqrt{A}\sqrt{1+f_{+}^{2} \left( \frac{x_{0}}{m_{0}}, \alpha
\right)}, \ \ \ \ E_{2}=m_{0}\alpha \varepsilon_{2} ,  \label{e2d}
\end{equation}
\begin{equation}
E_{1}=E_{0}-E_{2}=x_{1}\sqrt{A}=m_{0}g_{-} \left( \frac{x_{0}}{m_{0}},\alpha
\right) \sqrt{A} ,  \label{e1ddd}
\end{equation}
where $A$ is taken in the point of decay.

%%%%%%%%%%%%%%%%%%%%%%%%%%%%%%%%%%%%%%%%%%%%%%%%%%%%%%%%%%%%%%%%%%%%%%

\section{Example: escape from the ISCO}

    Let particle 0 rotate around a black hole on a circle orbit.
    Then, for a given $\mathcal{L} = L/m$, there are two radii, $r_{A}$
(stable) and $ r_{P}\leq r_{A}$ (unstable) \cite{LL},
\begin{equation}
\frac{r_{A}}{r_{g}}=\frac{\mathcal{L}^{2}}{r_{g}^{2}} \left[ 1 + \sqrt{1-
\frac{ 3r_{g}^{2}}{\mathcal{L}^{2}}} \right] ,  \label{ra}
\end{equation}
\begin{equation}
\frac{r_{P}}{r_{g}}=\frac{\mathcal{L}^{2}}{r_{g}^{2}} \left[ 1- \sqrt{1-
\frac{ 3r_{g}^{2}}{\mathcal{L}^{2}}} \right],  \label{rp}
\end{equation}
\begin{equation}
\varepsilon^{2}=U_{ef}(r_{c})=\frac{2A^{2}\mathcal{L}^{2}}{r_{g}r} ,
\label{eps}
\end{equation}
\begin{equation}
\mathcal{L}^{2}\geq 3r_{g}^{2} .
\end{equation}
In the case of equality, both roots coincide, so
\begin{equation}
\frac{r_{A}}{r_{g}}=3=\frac{r_{P}}{r_{g}}, \ \ \varepsilon^{2}=\frac{8}{9},
\ \ \frac{\mathcal{L}}{r_{g}}=\sqrt{3}, \ \ x_{A}\equiv \frac{m \mathcal{L}}{
r_{A}}=x_{P}=\frac{m}{\sqrt{3}}, \ \ \tilde{m}=\frac{2m}{\sqrt{3}},
\end{equation}
\begin{equation}
\frac{x_{A}}{m}=\frac{1}{\sqrt{3}}, \ \ \ \ \frac{\tilde{m}}{m}= \frac{2}{
\sqrt{3}},
\end{equation}
\begin{equation}
A=\frac{2}{3}, \ \ \ \ \sqrt{A}=\sqrt{\frac{2}{3}},
\end{equation}
\begin{equation}
\varepsilon =\frac{2\sqrt{2}}{3}=\sqrt{\frac{8}{9}} .
\end{equation}
    This corresponds to the innermost stable circular orbit (ISCO).

    In this case, eqs.~(\ref{f2}), (\ref{e2d}) give us
\begin{equation}
\varepsilon_{2}=\frac{3+\alpha^{2}}{3\sqrt{2}\alpha } .
\end{equation}
    Escape can occur if $\varepsilon \geq 1$.
    Then, $\alpha \leq \alpha_{-}$, where
\begin{equation}
\alpha_{-}=\frac{3-\sqrt{3}}{\sqrt{2}}<1 .
\end{equation}
In doing so, the velocity at infinity
\begin{equation}
v_{\infty }=\sqrt{1-\frac{1}{\varepsilon_{2}^{2}}}.
\end{equation}
If $\alpha \rightarrow \alpha_{-}$, $\varepsilon_{2}\rightarrow 1$,
\begin{equation}
\frac{E_{2}}{E_{0}}=\frac{3}{4}(3-\sqrt{3})\approx 0.951 ,
\end{equation}
\begin{equation}
\frac{E_{1}}{E_{0}}=\frac{3\sqrt{3}-5}{4}\approx 0.049 .
\end{equation}

%%%%%%%%%%%%%%%%%%%%%%%%%%%%%%%%%%%%%%%%%%%%%%%%%%%%%%%%%%%%%%%%%%%%%%

\section{Deceleration and departure from the vicinity of a horizon}

\hspace{0pt} The rocket stops in a point with the radial coordinate $r$,
provided $A(r)=\varepsilon ^{2}$. Let us consider the case of a photonic
rocket that halts after deceleration due to turning on the engine.
    Then, in~(\ref{fr}) $A(r)=\varepsilon _{2}^{2}$ and
    \begin{equation}
A=\varepsilon _{2}^{2},\ \ \ m_{2}=\frac{\varepsilon _{0}m_{0}}{\varepsilon
_{2}}\left( 1-\sqrt{1-\frac{A}{\varepsilon _{0}^{2}}}\right) .  \label{nk28}
\end{equation}
    Near the horizon, we have
\begin{equation}
\frac{A}{\varepsilon _{0}^{2}}\ll 1,\ \ m_{2}=\frac{\sqrt{A}m_{0}}{%
2\varepsilon _{0}}\left( 1+O\left( \frac{A}{\varepsilon _{0}^{2}}\right)
\right)   \label{nk29}
\end{equation}
    or
\begin{equation}
\frac{A}{\varepsilon _{0}^{2}}\ll 1,\ \ E_{2}=\frac{E_{0}}{2}\frac{A}{%
\varepsilon _{0}^{2}}\left( 1+O\left( \frac{A}{\varepsilon _{0}^{2}}\right)
\right) .  \label{nk30}
\end{equation}
    Thus if a rocket falls towards a black hole from a distance large as
compared the gravitational radius
$\left( \varepsilon _{0}\gg \sqrt{A(r)}\right)$,
it should radiate almost all its energy $E_{0}$ to stop.

Let now this rocket (having the mass $m_{2}$), after stopping, accelerates
due to fuel jet and departures away from a black hole achieving the initial
specific energy $\varepsilon _{0}$. According to (\ref{fr}), (\ref{nk28}),
we obtain the value $m_{3}$ for a rocket after this manoeuvre
    \begin{equation}
m_{3}=\frac{m_{2}\sqrt{A}}{\varepsilon _{0}\left( 1+\sqrt{1-\frac{A}{
\varepsilon _{0}^{2}}}\right) }=\frac{m_{0}A}{\varepsilon_{0}^{2}}\frac{1}{
\left( 1+\sqrt{1-\frac{A}{\varepsilon _{0}^{2}}}\right) ^{2}}.
\label{nk29dob}
\end{equation}
    In case of the Schwarzschild metric and $ \varepsilon_{0} =1$
Eq.~(\ref{nk29dob}) reproduces the result of problem (12-16) from~\cite{Hartle}.

    Near the horizon,
    \begin{equation}
\frac{A}{\varepsilon _{0}^{2}}\ll 1,\ \ \ \ m_{3}\approx \frac{m_{2}\sqrt{A}}{
2\varepsilon _{0}}\approx \frac{E_{0}A}{4\varepsilon _{0}^{3}}.
\label{nk29db}
\end{equation}

From eq. (\ref{fr}) with $\sigma _{v}=-1$ one can find the value of the
remaining mass. The result coincides with the expression given by the second
equality (\ref{nk29dob}).

    For a nonphotonic rocket decelerating to the state of rest, the relation
    \begin{equation}
\frac{\varepsilon _{0}m_{0}}{\varepsilon _{2}m_{2}}\left( \sqrt{1-
\frac{A}{ \varepsilon _{0}^{2}}}-\frac{|u|}{c}\right) =-\frac{|u|}{c}.
\label{nk31}
\end{equation}
should be fulfilled according to (\ref{prsi}). (We remind a reader that in
the deceleration regime, $u=-|u|$).
    The necessary condition for eq. (\ref{nk31}) to be valid, reads
\begin{equation}
\frac{|u|}{c}>\sqrt{1-\frac{A}{\varepsilon _{0}^{2}}}.  \label{nk32}
\end{equation}
    Near the horizon, when $A(r)/\varepsilon _{0}^{2}\ll 1$, it follows
from (\ref{nk32}) that the velocity of outflow of the jet fuel
should be close to that of light,
\begin{equation}
1-\frac{|u|}{c}<1-\sqrt{1-\frac{A}{\varepsilon _{0}^{2}}}\approx
\frac{A}{ 2\varepsilon _{0}^{2}}.  \label{nk32da}
\end{equation}

For the value of mass of a stopped rocket we have%
\begin{equation}
m_{2}=\frac{\varepsilon _{0}m_{0}}{\sqrt{A}}\left( 1-\frac{c}{|u|}
\sqrt{ 1 - \frac{A}{\varepsilon _{0}^{2}}}\right) .  \label{nk32do}
\end{equation}
    Obviously, $m_{2}$ changes from $m_{2}=0$ for a minimum possible
velocity $ |u|=c\sqrt{1-\frac{A}{\varepsilon _{0}^{2}}} $, necessary for
hovering, to a maximum value (\ref{nk28}) if an engine becomes photonic.

For acceleration of a rocket hovering over the horizon, to the initial value
$\varepsilon _{0}$, we obtain for a possible mass of ejected remnant
    \begin{equation}
m_{3}=\frac{m_{0}\left( 1-\frac{c}{|u|}\sqrt{1-\frac{A}{\varepsilon _{0}^{2}}
}\right) }{1+\frac{|u|}{c}\sqrt{1-\frac{A}{\varepsilon _{0}^{2}}}}\left( 1-
\frac{c}{|u|}\frac{\varepsilon _{0}^{2}}{A}\sqrt{1-\frac{A}{\varepsilon
_{0}^{2}}}\left( 1-\frac{u^{2}}{c^{2}}\right) \right) .  \label{nk32dop}
\end{equation}

    This equation applies to the two-step process described above.
    For the one-step scenario of the reverse of the velocity to the same
value $ \varepsilon _{0}$, we obtain for the remnant mass
\begin{equation}
m_{3}^{\ast }=m_{0}\frac{1-\frac{|u|}{c}\sqrt{1-\frac{A}{\varepsilon _{0}^{2}
}}-2\sqrt{1-\frac{A}{\varepsilon _{0}^{2}}}\frac{c}{|u|}\frac{\varepsilon
_{0}^{2}}{A}\left( 1-\frac{u^{2}}{c^{2}}\right) }{1+\frac{|u|}{c}\sqrt{1-
\frac{A}{\varepsilon _{0}^{2}}}}.  \label{nk32dpp}
\end{equation}

We have from~(\ref{nk32dop}), (\ref{nk32dpp})
\begin{equation}
m_{3}-m_{3}^{\ast }=m_{0}\left( \frac{\varepsilon _{0}^{2}}{A}-1\right)
\left( \frac{c^{2}}{u^{2}}-1\right) \geq 0.  \label{nk32srv}
\end{equation}
    Therefore, for a nonphotonic engine ($|u|<c$) the two-step process
(stopping and acceleration away from a black hole ) is more effective than
a one-step scenario.
    As a result, a larger mass can be taken out from the vicinity of
a horizon in such a process.

%%%%%%%%%%%%%%%%%%%%%%%%%%%%%%%%%%%%%%%%%%%%%%%%%%%%%%%%%%%%%%%%%%%%%%

\section{Continuous ejection and hovering over the horizon}

    Let us consider continuous process of fuel ejection.
    We assume that a body of mass $m$ ejects portion of fuel with the small
mass $dm^{\prime}$.
    Then, in the absence of external forces, the conservation law reads
\begin{equation}
D(mu^{\mu})+dm^{\prime}w^{\mu}=0 ,  \label{Dd}
\end{equation}
where $u^{\mu}$ corresponds to a rocket, $w^{\mu}$ corresponds to fuel. For
the flat space-time see discussion, e.g. in~\cite{henri}, pages~284--285.
    In~(\ref{Dd}) $D$ denotes covariant differential, so
    \begin{equation}
a^{\mu}=\frac{Du^{\mu}}{d\tau}=\frac{d^{2}x^{\mu}}{d\tau^{2}}+\Gamma
_{\alpha\beta}^{\mu}\frac{dx^{\alpha}}{d\tau}\frac{dx^{\beta}}{d\tau} ,
\label{a}
\end{equation}
    where $a^{\mu}$ is the four-acceleration, $\Gamma_{\alpha\beta}^{\mu}$ are
Christoffel symbols. Eq.~(\ref{Dd}) can be rewritten in the form
    \begin{equation}
dmu^{\mu}+ma^{\mu}d\tau+dm^{\prime}w^{\mu}=0 ,  \label{dm}
\end{equation}
$\tau$ is the proper time.

    Now, we are interested in the situation when a particle halts and hovers
over the horizon. Then, $u^{r}=0$.
    From $a^{\mu}u_{\mu}=0$ it follows that $ a^{0}=0$ as well.

As a result, we have from the $t-$component of~(\ref{dm}) that
\begin{equation}
dmu^{0}+dm^{\prime}w^{0}=0 .
\end{equation}
For particle at rest,
\begin{equation}
u^{0}=\frac{1}{\sqrt{A}} ,
\end{equation}
whence
\begin{equation}
dm^{\prime}=-\frac{dm}{w^{0}\sqrt{A}}.  \label{msh}
\end{equation}
From the radial component of~(\ref{dm}) we have
\begin{equation}
ma^{r}d\tau+dm^{\prime}w^{r}=0 .
\end{equation}
Here,
\begin{equation}
a^{r}=\Gamma_{00}^{r}\left( u^{0}\right)^{2}=\frac{A^{\prime}}{2} ,
\end{equation}
\begin{equation}
a^{2}=g_{\mu\nu}a^{\mu}a^{\nu}=\frac{A^{\prime2}}{4A} .  \label{a2}
\end{equation}
Combining these formulas, one obtains
\begin{equation}
\frac{dm}{d\tau}=m\frac{a^{r}\sqrt{A}}{w^{r}}w^{0} .  \label{mtmt}
\end{equation}

    Let we use tetrads $e_{(a)\mu}$, attached to a static observer,
so in the coordinates $(t,r,\theta,\phi)$
\begin{equation}
e_{(0)\mu}= \left( -\sqrt{A},0,0,0 \right) ,
\end{equation}
\begin{equation}
e_{(1)\mu}= \left( 0,\frac{1}{\sqrt{A}},0,0 \right) ,
\end{equation}
\begin{equation}
e_{(2)\mu}=r(0,0,1,0) ,
\end{equation}
\begin{equation}
e_{(2)\mu}=r\sin\theta(0,0,0,1) .
\end{equation}
Then,
\begin{equation}
w^{(1)}=\frac{1}{A}\frac{w^{r}}{w^{0}} ,
\end{equation}
whence in our case
\begin{equation}
\frac{w^{r}}{w^{0}}=-A\left\vert u\right\vert
\end{equation}
since eq.~(\ref{se3}) $v_{2}=0$ and $w^{r}<0$ in the deceleration regime.
Using also the normalization condition, one finds easily
\begin{equation}
w^{0}=\frac{\sqrt{\left( w^{r}\right)^{2}+A}}{A} ,
\end{equation}
\begin{equation}
w^{r}=-\frac{\left\vert u\right\vert \sqrt{A}}{\sqrt{1-u^{2}}} ,
\end{equation}
\begin{equation}
w^{0}=\frac{1}{\sqrt{A}}\frac{1}{\sqrt{1-u^{2}}} .
\end{equation}
Then, we have from~(\ref{mtmt}) for a given fixed~$r$
\begin{equation}
m=m_{0}\exp(-B\tau),  \label{mt}
\end{equation}
where
\begin{equation}
B=\frac{A^{\prime }}{2\sqrt{A}\left\vert u\right\vert },  \label{B}
\end{equation}
and we assumed that at $\tau =0$, $m=m_{0}$, $m^{\prime }=0$.
    This agrees with general formula~(5) of Ref.~\cite{HN12}.

    Then, taking into account~(\ref{msh}), we infer
    \begin{equation}
m^{\prime} = m_{0}\sqrt{1-u^{2}}[1-\exp (-B\tau )],  \label{f}
\end{equation}
    \begin{equation}
E_{\mathrm{loc}}= m_{0}[1-\exp (-B\tau )],  \label{loc}
\end{equation}
    where $E_{\mathrm{loc}}=m^{\prime} / \sqrt{1-u^{2}} $ is a local energy
measured by a stationary observer and expended by a rocket because of fuel
ejection. This quantity admits a safe limit to the case of a photon rocket,
when $u\rightarrow 1$, $m^{\prime }\rightarrow 0$ simultaneously,
$ E_{\mathrm{loc}} $ being finite.

    Meanwhile, there are serious restrictions on the possibility of the process
because of necessity to have a big initial mass.
    It is seen from eq.~(\ref{mt}) that fuel supply necessary for hovering
should be exponentially large as compared to the rocket mass.
    From~(\ref{mt}), we have the hovering proper time
\begin{equation}
\Delta \tau =\frac{1}{B}\log \frac{m_{0}}{m}  \label{vrzav}
\end{equation}
and the hovering time for a distant observer
\begin{equation}
\Delta t=\frac{1}{\sqrt{A}B}\log \frac{m_{0}}{m}  \label{vrtzav}
\end{equation}
where $m_{0}$ is initial mass the rocket with fuel and $m$ is final mass.

    In the Schwarzschild case we see that the rate with which mass decreases,
    \begin{equation}
-\frac{dm}{mdt}=\sqrt{A}B=\frac{r_{+}c^{2}}{2r^{2}|u|}  \label{fsk}
\end{equation}
    remains finite even near the horizon $ r_{+} $ and changes smoothly when
a point of hovering becomes closer and closer to the horizon.
    Meanwhile, for a local observer, the corresponding rate is
    \begin{equation}
-\frac{dm}{md\tau }=B=\frac{r_{+}c^{2}}{2r^{2}|u|\sqrt{1-\frac{r+}{r}}}.
\label{fskl}
\end{equation}
    In the horizon limit, because of the redshift factor $\sqrt{A}$
in~(\ref{B}), this rate changes crucially depending on a point and diverges
when the horizon is approached.

    To illustrate the situation by concrete examples, we assume that a rocket
with a photon engine hovers over the horizon of a black hole that has the
Sun mass and the initial mass of a rocket is of the order $10^{12}$\,kg.
    For substance with usual density the size of such a spacecraft has
the order $1$\,km.
    Then, after the time of the order $10^{-3}$\,s the remaining mass of a
rocket with fuel cannot exceed the mass of the atom of
hydrogen $ 1.67 \cdot 10^{-27}$\,kg.

    In another example, a photon super-rocket with the mass equal to that of
Earth ($5.97\cdot 10^{24}$\,kg) hovers over a black hole in the center of
Milky Way.
    We assume that a black hole has the mass equal to $4.3\cdot
10^{6} $ masses of Sun. Then, after one and a half hours of hovering the
mass of a rocket with remnants of fuel cannot exceed $100$\,kg.

%%%%%%%%%%%%%%%%%%%%%%%%%%%%%%%%%%%%%%%%%%%%%%%%%%%%%%%%%%%%%%%%%%%%%%

\section{Conclusion}

    We studied the Oberth effect in the relativistic case. In particular, we
showed that this effect enables us to convert the total energy ($E=mc^{2}$)
of jet fuel into the kinetic energy of a photon rocket, provided the process
occurs near the black hole horizon of the Schwarzschild case. In this sense,
the relativistic Oberth effect for nonrotating black holes is very close to
the Penrose process, when decay of a particle inside the ergosphere of a
rotating black hole produces debris whose energy measured at infinity
exceeds the initial energy. It is possible when one of new particles remains
on an orbit with a negative energy. As for a nonrotating neutral black hole
the negative energy states are absent, the Penrose effect is absent as well.
However, near the horizon there are states with the energy whose value is as
small as one likes. As a result, after decay the energy of one of fragments
can be close to the energy of an initial particle as close as one likes. For
the evaluation of role of the Oberth effect we introduced the efficiency in
the relativistic case and discussed its behavior in some typical cases. We
directly proved that the ejection of fuel along the trajectory corresponds
to the most efficient scenario for given initial energy, angular momentum
and fixed masses of particles participating in decay. We also considered the
process in which a rocket continuously ejects fuel in such a way that this
enables it to hover over a horizon. Then, the consumption of fuel should
change exponentially with respect to the proper time.

It is worth noting that sometimes an oversimplified interpretation of the
Oberth effect is used that does not show the essence of matter properly. For
example, in the paper ``Oberth effect'' from Wikipedia it is said that ``the
Oberth effect, wherein the use of a reaction engine at higher speeds
generates a greater change in mechanical energy than its use at lower
speeds''. This does not into account a simple circumstance: we must at first
drive a rocket at high speed starting from low velocities. Then, on the
first stage of the process an ineffective expenses of fuel are inevitable.
The idea suggested by Oberth consisted in the gain of high velocities due to
the action of the gravitational field. Then, the use of an engine in the
periastron of a trajectory becomes effective~\cite{bm2}.
    In modern astronautics, the Oberth effect is used for gravitational
maneuvers~\cite{ad}.
    In corresponding situations, the relativistic effects are small or even
negligible. Meanwhile, in our paper, we showed that the efficiency of the
acceleration of the jet engine can reach 100\,\% on the event horizon. Then,
the whole energy (including the rest one) is spent to the growth of the
rocket energy.

    We would also like to stress the following difference between the
nonrelativistic and relativistic versions of the Oberth effect. In the first
case, we saw that, according to~(\ref{ob}), the efficiency admits splitting
to the contribution due to the energy stored in fuel and kinematic
correction (the Oberth effect is described just the second term giving an
additive correction to the first term). However, in the relativistic case
(when velocities are high and the gravitational field is strong) such
decomposition is not valid and the manifestation of the Oberth effect is
essentially nonlinear.

    We hope that consideration of the simplest case of the static spherically
symmetric black hole will be useful for further generalization to rocket
movement in the vicinity of rotating black holes.

\vspace{11pt}
%%%%%%%%%%%%%%%%%%%%%%%%%%%%%%%%%%%%%%%%%%%%%%%%%%%%%%%%%%%%%%%%%%%%%%
\textbf{Acknowledgements.} \ The work of Yu.V.P. has been supported by the
Kazan Federal University Strategic Academic Leadership Program and the
Russian Scientific Foundation (Grant No. 22-22-00112).

%%%%%%%%%%%%%%%%%%%%%%%%%%%%%%%%%%%%%%%%%%%%%%%%%%%%%%%%%%%%%%%%%%%%%%

\end{document}